\documentclass[a4paper,12pt]{article}
\pdfoutput=1
\usepackage{amssymb}
\usepackage{tipa}
\usepackage{mathrsfs}
\usepackage{bbm}
\usepackage{epsf,amsmath}
\usepackage[dvips,usenames]{color}
\usepackage{graphicx,subfigure}
\usepackage{xcolor}
\usepackage{colortbl}
\usepackage{booktabs,multirow,makecell}
\usepackage{float}
\usepackage{threeparttable}
\usepackage{cite}
\usepackage{hyperref}  
 \allowdisplaybreaks[4]
\newlength{\dinwidth}
\newlength{\dinmargin}
\def\be{\begin{equation}}
\def\ee{\end{equation}}
\def\ba{\begin{eqnarray}}
\def\ea{\end{eqnarray}}

\def\be{\begin{equation}}
\def\ee{\end{equation}}
\def\ba{\begin{eqnarray}}
\def\ea{\end{eqnarray}}

        \def \d {{\rm d}}

          \def\qb{{ \bf q}_\bot}
           
           \def\kb{{ \bf k}_\bot}

\begin{document}
\title{\bf The purely leptonic and semileptonic decays of $D^{*}_{s}$ meson}
\author{Xiao-Lin Wang$^{a}$\href{https://orcid.org/0009-0003-3016-8082}{\includegraphics[scale=0.3]{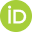}}, Rui Sun$^{a}$, Li-Ting Wang$^{a}$\footnote{wanglt@htu.edu.cn}~\href{https://orcid.org/0009-0004-1725-5525}{\includegraphics[scale=0.3]{ORCID.png}} \\
{ $^a$\small Centre for Theoretical Physics, Henan Normal University, Henan 453007, China}}
\date{}
\maketitle

\begin{abstract}
 With the potential prospects of the $D^{*}_{s}$ at high-luminosity heavy-flavor experiments in the future, we investigated the CKM-favored and tree-dominated leptonic $D^{*}_{s}\to\ell\bar{\nu}_{\ell}$ and semileptonic $D^{*}_{s}\to M\ell\bar{\nu}_{\ell}$ ($M=\phi, \eta^{(\prime)}$ and $\ell=e, \mu$) weak decays in the Standard Model (SM). The theoretical predictions and some discussions for the observable quantities including the total width of $D^{*}_{s}$ mesons, the branching fractions of leptonic $D^{*}_{s}\to\ell\bar{\nu}_{\ell}$ and semileptonic $D^{*}_{s}\to M\ell\bar{\nu}_{\ell}$ weak decays, the lepton spin asymmetry and forward-backward asymmetry are presented. Numerically, the weak decays of $D^{*}_{s}\to\ell\bar{\nu}_{\ell}$ and $D^{*}_{s} \to M\ell\bar{\nu}_{\ell}$ have relatively large branching fractions of the order $\mathcal{O}(10^{-5})$ and $\mathcal{O}(10^{-7})$ respectively, which are expected to be observed in future experiments.\\
\end{abstract}

\newpage
\section{Introduction}
Since  the SPEAR experiment found the existence of charmed mesons in positron-electron annihilation in 1976~\cite{Goldhaber:1976xn,Peruzzi:1976sv}, its production and decay properties have been one of the important topics in particle physics.
The pseudoscalar $D_{s}$ meson  and vector $D^{*}_{s}$ meson are consist of a quark-antiquark pair $c\bar{s}$, and have the same  quantum numbers of electric charge,  charm and strange, i.e., $C=-S=Q=+1$. The different spin configurations of interquark potential make the mass of the ground spin-triplet $1^{3}S_{1}$ state for the $D^{*}_{s}$ meson to be above that of the ground spin-singlet $1^{1}S_{0}$ state for the $D_{s}$ meson~\cite{PDG:2022}. The weak decays of $D_{s}$ meson have been widely studied in the past decades, while  there are few theoretical studies of $D^{*}_{s}$  weak decays  even though it is legal and allowable within the Standard Model~(SM), because the $D_s^*$ meson decay occur mainly through radiative processes $D^{*}_{s}\to D_s \gamma$,  and its weak decays are generally too rare to be measured by previous experiments.

Recently, BESIII collaboration reports the first experimental study of the CKM-favored purely leptonic weak decay $D^{*+}_{s}\to e^+\nu_e$. Its branching fraction is measured
to be~\cite{BESIII:2023zjq}
\begin{align}
\mathcal{B}(D^{*+}_{s}\to e^+\nu_e) = (2.1^{+1.2}_{-0.9}\pm0.2)\times 10^{-5},\qquad \text{BESIII  }
\label{eq:bes}
\end{align}
corresponding to an upper limit of $4.0\times 10^{-5}$ at the $90\%$ confidence level.
With the continuous upgrading of experimental technology and equipment, more data on $D^{*}_{s}$ mesons is expected to be collected in future high energy physical experiments. For example, in the $e^{+}e^{-}$ colliders, it is promisingly expected that there will be a total of about $5\times10^{10}$ $c\bar{c}$ pairs at the SuperKEKB\cite{EPJCSun:10}. Considering the fraction of the charmed quark fragmenting into the $D^{*}_{s}$ meson $f(c\to D^{*}_{s})\simeq 5.5\%$\cite{EPJCSun:13}, the high statistical $c\bar{c}$ pairs correspond to about $6\times 10^{9}$ $D^{*}_{s}$ mesons at the SuperKEKB. In the high energy hadron colliders, about $4\times 10^{13}$ $D^{*}_{s}$ mesons\cite{EPJCSun:14} are expected to be collected with a data sample of target luminosity $300fb^{-1}$ at the LHCb@HL-LHC experiments\cite{EPJCSun:17}. Besides, the experiments at upcoming Super Tau-Charm Facility~(STCF) and the Circular Electron Positron Collider (CEPC) can also provide  a large amount of information for the $D^{*}_{s}$  weak decays.

The $D^{*}_{s}$ meson decay is dominated by the radiative process, while it can also occur through weak interactions within the SM. As is well known, the strong and electromagnetic interactions are related only to vector currents, while the weak interaction is not only related  to vector currents, but also is related to axial vector currents. In this paper, we would like to study the purely leptonic and semileptonic decays of $D^{*}_{s}$ meson, which can not only enhance our understanding of the properties $D^{*}_{s}$ mesons, but also provide a stringent test of the axial vector current interactions. In recent years, some hints of lepton flavor universality violation in semileptonic $B$ decays have been reported by a lot of experimental collaborations~\cite{PRL127:3,PRL127:4,PRL127:5,PRL127:6,PRL127:7} and are widely studied by theorists~\cite{Martinelli:2022fgg,Sheng:2022peg,Chang:2018sud,Bigi:2016mdz}. It is argued that the violation may also occur in $c\to s\ell\bar{\nu}_{\ell}$ transition because the mechanism of the $c\to s\ell\bar{\nu}_{\ell}$ transition is similar to the one of $b\to c\ell\bar{\nu}_{\ell}$ transition.
Therefore, the phenomenological study of the semileptonic $D^{*}_{s}\to M\ell\bar{\nu}_{\ell}$ ($M=\phi, \eta^{(\prime)}$) weak decays may provide some valuable references for testing the lepton flavor universality and the relevant measurement in the future experiments. Explicitly, in this paper, we will pay our attention to the CKM-favored and tree-dominated leptonic $D^{*}_{s}\to \ell\bar{\nu}_{\ell}$ and semileptonic $D^{*}_{s}\to M\ell\bar{\nu}_{\ell}$ ($M=\phi, \eta^{(\prime)}$) weak decays, which are generally much more complicated than the corresponding $D$ decay modes because they involves much more allowed helicity states.

Our paper is organized as follows. In Section 2, the helicity amplitudes and observables of  $D^{*}_{s}\to\ell\bar{\nu}_{\ell}$ and $D^{*}_{s}\to M\ell\bar{\nu}_{\ell}$  decays are calculated. Section 3 is devoted to the numerical results and discussions. In the calculation, the  form factors of $D^{*}_{s}\to M$ transition are obtained by using the self-consistent covariant light-front quark model. Finally, we give our summary in Section 4.
\section{Theoretical framework}
In the SM, $D^{*}_{s}\to\ell\bar{\nu}_{\ell}$ and $M\ell\bar{\nu}_{\ell}$ decays are induced by $c \to s\ell\bar{\nu}_\ell$ transition at quark level via W-exchange, and can be described by the effective Hamiltonian
\begin{align}
\mathcal{H}_{\rm eff}(c \to s\ell\bar{\nu}_\ell) &= \frac{G_{F}}{\sqrt{2}}V_{cs}
[\bar{s}\gamma_{\mu}(1-\gamma_{5})c][\bar{\ell}\gamma^{\mu}(1-\gamma_{5})\nu_{\ell}]\,,
\label{eq:Heff}
\end{align}
where $G_{F}$ is the Fermi constant and $V_{cs}$ is the CKM matrix element.

Using Eq.~\eqref{eq:Heff}, the amplitude of  $D^{*}_{s}\to\ell\bar{\nu}_{\ell}$ decay can be written as,
\begin{align}
 \langle \ell\bar{\nu}_\ell | \mathcal{H}_{\rm eff}  | D^{*}_{s}  \rangle = \frac{G_{F}}{\sqrt{2}}V_{cs} \langle \ell\bar{\nu}_\ell | \gamma^{\mu}(1-\gamma_{5}) | 0  \rangle \, \langle 0 | \gamma_{\mu}(1-\gamma_{5}) | D^{*}_{s}  \rangle \,.
\end{align}
The leptonic part can be calculated reliably within perturbative theory and the
hadronic matrix element can be parameterized in terms of decay constants.
Then, the helicity amplitudes can be easily obtained and are written as,
\begin{align}
|M(D^{*}_{s}\to\ell\bar{\nu}_{\ell})_{\lambda_{\ell}=1/2}|^{2}&=\frac{G_{F}^{2}|V_{cs}|^{2}}{2}
\frac{1}{2}m^{2}_{\ell}\left(m^{2}_{D^{*}_{s}}-m^{2}_{\ell}\right)f^{2}_{D^{*}_{s}}\,,
\label{eq:lepM1}\\
|M(D^{*}_{s}\to\ell\bar{\nu}_{\ell})_{\lambda_{\ell}=-1/2}|^{2}&=\frac{G_{F}^{2}|V_{cs}|^{2}}{2}
m^{2}_{D^{*}_{s}}\left(m^{2}_{D^{*}_{s}}-m^{2}_{\ell}\right)f^{2}_{D^{*}_{s}}\,.
\label{eq:lepM2}
\end{align}
Comparing Eq.~\eqref{eq:lepM1} with Eq.~\eqref{eq:lepM2}, one can be obviously find that $D^{*}_{s}\to\ell\bar{\nu}_{\ell}$ is dominated by the $\lambda_{\ell}=-1/2$ state, and the amplitudes of $\lambda_{\ell}=1/2$ state is suppressed by $m_{\ell}/m_{D^{*}_{s}}$. Using above amplitudes, one can finally obtained the decay width, which is written as~\cite{EPJCSun:14},
\begin{align}
\Gamma({D^{*}_{s}\to \ell\bar{\nu}_{\ell}})=\frac{G_{F}^2}{12 \pi } |V_{cs}|^{2} f^{2}_{D^{*}_{s}} m^{3}_{D^{*}_{s}}\left(1-\frac{m^{2}_{\ell}} {m^{2}_{D^{*}_{s}}}\right)^2\left(1+\frac{m^{2}_{\ell}} {2 m^{2}_{D^{*}_{s}} }\right)\,.
\end{align}

For the semileptonic $D^{*}_{s} \to M\ell\bar{\nu}_{\ell}$ decays, the calculations of amplitudes are relatively complicated.
Using Eq. (\ref{eq:Heff}), the square amplitude can be expressed as
\begin{align}
|{\mathcal{M}(D^{*}_{s} \to M\ell\bar{\nu}_{\ell})}|^{2} &=\frac{G_{F}^{2}{|V_{cs}|^{2}}}{2} |\langle{M|\bar{s}\gamma_{\mu}(1-\gamma_{5})c|D^{*}_{s}}\rangle\bar{\ell}\gamma^{\mu}
(1-\gamma_{5})\nu_{\ell}|^{2}\nonumber\\
& \equiv \frac{G_{F}^{2}{|V_{cs}|^{2}}}{2}L_{\mu\nu}H^{\mu\nu},
\end{align}
where $L_{\mu\nu}$ and $H^{\mu\nu}$ are leptonic and hadronic tensors built from the respective products of the lepton and hadron currents. After inserting the completeness relation of the polarization vectors of virtual $W$ boson,
$\sum_{m,n}\bar{\epsilon}_{\mu}(m)\bar{\epsilon}^{*}_{\nu}(n)g_{mn}=g_{\mu\nu}$,
the product of $L_{\mu\nu}$ and $H^{\mu\nu}$ can be expressed as
\begin{align}
L_{\mu\nu}H^{\mu\nu} &= \sum_{m,m',n,n'}L(m,n)H(m',n')g_{mm'}g_{nn'}\label{eq:L},
\end{align}
with
\begin{align}
L(m,n)  \equiv L^{\mu\nu}\bar{\epsilon}_{\mu}(m)\bar{\epsilon}^{*}_{\nu}(n), \quad
H(m',n') \equiv H^{\mu\nu}\bar{\epsilon}^{*}_{\mu}(m')\bar{\epsilon}_{\nu}(n'),
\end{align}
where $\bar{\epsilon}_{\mu} (m)$ with $m=t, 0, \pm1$ is the polarization vector of $W$ and ${g}_{mm'}=\rm{diag(+1,-1,-1,-1)}$. A benefit of expression of Eq.(\ref{eq:L}) is that $H(m',n')$ and $L(m,n)$ are Lorentz invariant and therefore can be evaluated in different reference frames~\cite{Korner:1987kd,Korner:1989qb,Hagiwara:1989cu,Hagiwara:1989gza}. For convenience, $H(m',n')$ and $L(m,n)$ are generally calculated in the rest frame of $D^{*}_{s}$ meson and the $\ell-\bar{\nu}_{\ell}$ center-of-mass frame, respectively.

For the hadronic part, one has to calculate the hadronic helicity amplitude $H_{\lambda_{W}\lambda_{D^{*}_{s}}\lambda_{M}}$ of $D^{*}_{s} \to M\ell\bar{\nu}_{\ell}$ decays, defined by
\begin{align}
H_{\lambda_{W}\lambda_{D^{*}_{s}}\lambda_{M}}(q^{2}) &= H_{\mu}(\lambda_{D^{*}_{s}},\lambda_{M})\bar{\epsilon}^{*\mu}(\lambda_{W})\nonumber\\
&=\langle{M(\lambda_{M},p_{M})|\bar{s}\gamma_{\mu}(1-\gamma_{5})c|D^{*}_{s}(\lambda_{D^{*}_{s}},p_{D^{*}_{s}})}\rangle
\bar{\epsilon}^{*\mu}(\lambda_{W})\,,
\end{align}
which describes the decay of three helicity states of $D^{*}_{s}$ meson into a pseudo-scalar $\eta^{(\prime)}$ meson or the three helicity states of daughter $\phi$-meson and the four helicity states of $W$ boson. The quantity $H_{\mu}(\lambda_{D^{*}_{s}},\lambda_{M})$ represents the hadronic matrix elements of $D^{*}_{s} \to M$ transition, which can be parameterized in terms of form factors $V_{i}(q^{2})$ and $A_{i}(q^{2})$ as \cite{V2Vlnu:81,V2Vlnu:82,V2Plnu:QinChang},
\begin{align}
\langle{\phi(\epsilon_{2},p_{\phi})|\bar{s}\gamma_{\mu}c|D^{*}_{s}(\epsilon_{1},p_{D^{*}_{s}})}\rangle
=&(\epsilon_{1}\cdot\epsilon^{*}_{2})[-P_{\mu}V_{1}(q^{2})+q_{\mu}V_{2}(q^{2})] \nonumber\\
&+\frac{(\epsilon_{1}\cdot{q})(\epsilon^{*}_{2}\cdot{q})}{m^{2}_{D^{*}_{s}}-m^{2}_{\phi}}
[P_{\mu}V_{3}(q^{2})-q_{\mu}V_{4}(q^{2})] \nonumber\\
&-(\epsilon_{1}\cdot{q})\epsilon^{*}_{2,\mu}V_{5}(q^{2})+(\epsilon^{*}_{2}\cdot{q})\epsilon_{1,\mu}V_{6}(q^{2})\,,
\label{eq:FF1} \\
\langle{\phi(\epsilon_{2},p_{\phi})|\bar{s}\gamma_{5}\gamma_{\mu}c|D^{*}_{s}(\epsilon_{1},p_{D^{*}_{s}})}\rangle
=&-i\varepsilon_{\mu\nu\alpha\beta}\epsilon^{\alpha}_{1}\epsilon^{*\beta}_{2}
[P^{\nu}A_{1}(q^{2})-q^{\nu}A_{2}(q^{2})]\nonumber\\
&-\frac{i\epsilon^{*}_{2}\cdot{q}}{m^{2}_{D^{*}_{s}}-m^{2}_{\phi}}\varepsilon_{\mu\nu\alpha\beta}\epsilon^{\nu}_{1}
P^{\alpha}q^{\beta}A_{3}(q^{2})\nonumber\\
&+\frac{i\epsilon_{1}\cdot{q}}{m^{2}_{D^{*}_{s}}-m^{2}_{\phi}}\varepsilon_{\mu\nu\alpha\beta}\epsilon^{*\nu}_{2}
P^{\alpha}q^{\beta}A_{4}(q^{2})\,,
\label{eq:FF2}\\
\langle{\eta^{(\prime)}(p_{\eta^{(\prime)}})|\bar{s}\gamma_{\mu}c|D^{*}_{s}(\epsilon_{1},p_{D^{*}_{s}})}\rangle
=&-\frac{2iV(q^{2})}{m_{D^{*}_{s}}+m_{\eta^{(\prime)}}}\varepsilon_{\mu\nu\rho\sigma}\epsilon^{\nu}p^{\rho}_{\eta^{(\prime)}}p^{\sigma}_{  D^{*}_{s}}\,,
\label{eq:FF3}\\
\langle{\eta^{(\prime)}(p_{\eta^{(\prime)}})|\bar{s}\gamma_{\mu}\gamma_{5}c|D^{*}_{s}(\epsilon_{1},p_{D^{*}_{s}})}\rangle
=&2m_{D^{*}_{s}}A_{0}(q^{2})\frac{\epsilon_{1}\cdot q}{q^{2}}q^{\mu}
+(m_{\eta^{(\prime)}}+m_{D^{*}_{s}})A_{1}(q^{2})\left(\epsilon_{\mu}-\frac{\epsilon_{1}\cdot q}{q^{2}}q_{\mu}\right)\nonumber\\
&+A_{2}(q^{2})\frac{\epsilon_{1}\cdot q}{m_{\eta^{(\prime)}}+m_{D^{*}_{s}}}\left[P_{\mu}-\frac{m^{2}_{D^{*}_{s}}-m^{2}_{\eta^{(\prime)}}}{q^{2}}q_{\mu}\right]\,,
\label{eq:FF4}
\end{align}
with $P= p_{D^{*}_{s}}+p_{M}$, $q= p_{D^{*}_{s}}-p_{M}$ and the sign convention $\epsilon_{0123}=-1$.
Then, by contracting these hadronic matrix elements with the polarization vectors of virtual $W^{*}$ boson, one can finally obtain non-vanishing hadronic helicity amplitudes of $D^{*}_{s} \to M \ell \bar{\nu}_{\ell}$ decays.
For $D^{*}_{s} \to \eta^{(\prime)} \ell\bar{\nu}_{\ell}$ decay, the helicity amplitudes $H_{\lambda_{W}\lambda_{D^{*}_{s}}}$ are given by,
\begin{align}
H_{t0}(q^{2})=&-\frac{2m_{D^{*}_{s}}|\vec{p}_{\eta^{(\prime)}}|}{\sqrt{q^{2}}}A_{0}(q^{2})\,,
\label{eq:HAetas1}\\
H_{00}(q^{2})=&-\frac{1}{2m_{D^{*}_{s}}\sqrt{q^{2}}}\left[(m_{D^{*}_{s}}+m_{\eta^{(\prime)}})(m^{2}_{D^{*}_{s}}-m^{2}_{\eta^{(\prime)}}+ q^{2})A_{1}(q^{2})+\frac{4m^{2}_{D^{*}_{s}}|\vec{p}_{\eta^{(\prime)}}|^{2}}{m_{D^{*}_{s}}+m_{\eta^{(\prime)}}}A_{2}(q^{2})\right]\,,\\
H_{\mp\pm}(q^{2})=&(m_{D^{*}_{s}}+m_{\eta^{(\prime)}})A_{1}(q^{2})\pm\frac{2m_{D^{*}_{s}}|\vec{p}_{\eta^{(\prime)}}|}{m_{D^{*}_{s}}+m_{\eta^{(\prime)}}}V(q^{2})\,. \label{eq:HAetas3}
\end{align}
It is crucial to account for $\eta$-$\eta'$ mixing, which connects the physical states to the flavor eigenstates $\eta_q (q=u, d)$ and $\eta_s$ through
\begin{equation*}
\left(\begin{array}{c}
\eta\\
\eta^{\prime}
\end{array}\right)
=
\left(\begin{array}{cc}
\textrm{cos}\phi  & -\textrm{sin}\phi \\
\textrm{sin}\phi  & \textrm{cos}\phi
\end{array}\right)
\left(\begin{array}{c}
\eta_{q} \\
\eta_{s}
\end{array}\right)\,,
\end{equation*}
where $\phi$ is the mixing angle and has been well determined as $\phi=40^{\circ}$\cite{BESIII:2023ajr}. Since the $D_s^{*}$ couples primarily to the strange component, the physical helicity amplitudes for $D_s^{*} \to \eta^{(\prime)} \ell\bar{\nu}_{\ell}$ are accordingly obtained from the corresponding $\eta_s$ amplitudes.

For $D^{*}_{s} \to \phi \ell \bar{\nu}_{\ell}$ decay, the hadronic helicity amplitudes $H_{\lambda_{W}\lambda_{D^{*}_{s}}\lambda_{\phi}}$ are given as
\begin{align}
H_{0++}(q^{2})=&-\frac{m^{2}_{D^{*}_{s}}-m^{2}_{\phi}}{\sqrt{q^{2}}}A_{1}(q^{2})
+\sqrt{q^{2}}A_{2}(q^{2})+\frac{2m_{D^{*}_{s}}|\vec{p}_{\phi}|}{\sqrt{q^{2}}}V_{1}(q^{2}),\\
H_{t++}(q^{2})=&-\frac{2m_{D^{*}_{s}}|\vec{p}_{\phi}|}{\sqrt{q^{2}}}A_{1}(q^{2})
+\frac{m^{2}_{D^{*}_{s}}-m^{2}_{\phi}}{\sqrt{q^{2}}}V_{1}(q^{2})-\sqrt{q^{2}}V_{2}(q^{2}),\\
H_{-+0}(q^{2})=&-\frac{m^{2}_{D^{*}_{s}}+3m^{2}_{\phi}-q^{2}}{2m_{\phi}}A_{1}(q^{2})
+\frac{m^{2}_{D^{*}_{s}}-m^{2}_{\phi}-q^{2}}{2m_{\phi}}A_{2}(q^{2})\nonumber\\
&-\frac{2m^{2}_{D^{*}_{s}}|\vec{p}_{\phi}|^{2}}{m_{\phi}(m^{2}_{D^{*}_{s}}-m^{2}_{\phi})}A_{3}(q^{2})
-\frac{m_{D^{*}_{s}}|\vec{p}_{\phi}|}{m_{\phi}}V_{6}(q^{2}),\\
H_{0--}(q^{2})=&\frac{m^{2}_{D^{*}_{s}}-m^{2}_{\phi}}{\sqrt{q^{2}}}A_{1}(q^{2})
-\sqrt{q^{2}}A_{2}(q^{2})
+\frac{2m_{D^{*}_{s}}|\vec{p}_{\phi}|}{\sqrt{q^{2}}}V_{1}(q^{2}),\\
H_{t--}(q^{2}) =&\frac{2m_{D^{*}_{s}}|\vec{p}_{\phi}|}{\sqrt{q^{2}}}A_{1}(q^{2})
+\frac{m^{2}_{D^{*}_{s}}-m^{2}_{\phi}}{\sqrt{q^{2}}}V_{1}(q^{2})-\sqrt{q^{2}}V_{2}(q^{2}),\\
H_{+-0}(q^{2}) =&
\frac{m^{2}_{D^{*}_{s}}+3m^{2}_{\phi}-q^{2}}{2m_{\phi}}A_{1}(q^{2})
-\frac{(m^{2}_{D^{*}_{s}}-m^{2}_{\phi}-q^{2})}{2m_{\phi}}A_{2}(q^{2}) \nonumber\\
&+\frac{2m^{2}_{D^{*}_{s}}|\vec{p}_{\phi}|^{2}}{m_{\phi}(m^{2}_{D^{*}_{s}}-m^{2}_{\phi})}A_{3}(q^{2})
-\frac{m_{D^{*}_{s}}|\vec{p}_{\phi}|}{m_{\phi}}V_{6}(q^{2}),\\
H_{+0+}(q^{2})=&\frac{3m^{2}_{D^{*}_{s}}+m^{2}_{\phi}-q^{2}}{2m_{D^{*}_{s}}}A_{1}(q^{2})
-\frac{(m^{2}_{D^{*}_{s}}-m^{2}_{\phi}+q^{2})}{2m_{D^{*}_{s}}}A_{2}(q^{2})\nonumber\\
&+\frac{2m_{D^{*}_{s}}|\vec{p}_{\phi}|^{2}}{(m^{2}_{D^{*}_{s}}-m^{2}_{\phi})}A_{4}(q^{2})
-|\vec{p}_{\phi}|V_{5}(q^{2}),\\
H_{-0-}(q^{2})=&-\frac{(3m^{2}_{D^{*}_{s}}+m^{2}_{\phi}-q^{2})}{2m_{D^{*}_{s}}}A_{1}(q^{2})
+\frac{m^{2}_{D^{*}_{s}}-m^{2}_{\phi}+q^{2}}{2m_{D^{*}_{s}}}A_{2}(q^{2})\nonumber\\
&-\frac{2m_{D^{*}_{s}}|\vec{p}_{\phi}|^{2}}{m^{2}_{D^{*}_{s}}-m^{2}_{\phi}}A_{4}(q^{2})
-|\vec{p}_{\phi}|V_{5}(q^{2}),\\
H_{000}(q^{2})=&\frac{(m^{2}_{D^{*}_{s}}+m^{2}_{\phi}-q^{2})|\vec{p}_{\phi}|}{m_{\phi}\sqrt{q^{2}}}V_{1}(q^{2})
+\frac{2m^{2}_{D^{*}_{s}}|\vec{p}_{\phi}|^3}{m_{\phi}\sqrt{q^{2}}(m^{2}_{D^{*}_{s}}-m^{2}_{\phi})}V_{3}(q^{2})\nonumber\\
&-\frac{(m^{2}_{D^{*}_{s}}-m^{2}_{\phi}-q^{2})|\vec{p}_{\phi}|}{2m_{\phi}\sqrt{q^{2}}}V_{5}(q^{2})
+\frac{(m^{2}_{D^{*}_{s}}-m^{2}_{\phi}+q^{2})|\vec{p}_{\phi}|}{2m_{\phi}\sqrt{q^{2}}}V_{6}(q^{2}),\\
H_{t00}(q^{2})=&\frac{(m^{2}_{D^{*}_{s}}+m^{2}_{\phi}-q^{2})(m^{2}_{D^{*}_{s}}-m^{2}_{\phi})}{2m_{D^{*}_{s}}m_{\phi}\sqrt{q^{2}}}V_{1}(q^{2})
-\frac{(m^{2}_{D^{*}_{s}}+m^{2}_{\phi}-q^{2})\sqrt{q^{2}}}{2m_{D^{*}_{s}}m_{\phi}}V_{2}(q^{2})\nonumber\\
&+\frac{m_{D^{*}_{s}}|\vec{p}_{\phi}|^{2}}{m_{\phi}\sqrt{q^{2}}}V_{3}(q^{2})
-\frac{m_{D^{*}_{s}}|\vec{p}_{\phi}|^{2}\sqrt{q^{2}}}{m_{\phi}(m^{2}_{D^{*}_{s}}-m^{2}_{\phi})}V_{4}(q^{2})
-\frac{m_{D^{*}_{s}}|\vec{p}_{\phi}|^{2}}{\sqrt{q^{2}}m_{\phi}}V_{5}(q^{2})+\frac{m_{D^{*}_{s}}|\vec{p}_{\phi}|^{2}}{\sqrt{q^{2}}m_{\phi}}V_{6}(q^{2})\,,
\end{align}
in which, $\lambda_{W}=t$ has to be understood as $\lambda_{W}=0$ with $J=0$. It is obvious that only the amplitudes with $\lambda_{D^{*}_{s}}=\lambda_{\eta^{(\prime)}}-\lambda_{W}=-\lambda_{W}$ and  $\lambda_{D^{*}_{s}}=\lambda_{\phi}-\lambda_{W}$ survive due to the helicity conservation.

For the leptonic part, the leptonic tensor is usually expanded in terms of a complete set of Wigner's $d^{J}$-functions~\cite{Korner:1987kd,Fajfer:2012vx,Kadeer:2005aq}. As a result, the product $L_{\mu\nu}H^{\mu\nu}$ can be reduced to a compact form
\begin{align}
L_{\mu\nu}H^{\mu\nu}=&\frac{1}{8}\sum_{\lambda_{\ell},\lambda_{\bar{\nu}_{\ell}},
J,J' ,\lambda_{W}\lambda_{W^{\prime}}}
(-1)^{J+J'}|h_{\lambda_{\ell},\lambda_{\bar{\nu_{\ell}}}}|^{2}
\delta_{\lambda_{D^{*}_{s}},\lambda_{M}-\lambda_{W}}\delta_{\lambda_{D^{*}_{s}},\lambda_{M}-\lambda'_{W}}\nonumber\\
&\times{d^{J}_{\lambda_{W},\lambda_{\ell}-\frac{1}{2}}d^{J'}_{\lambda'_{W},\lambda_{\ell}-\frac{1}{2}}}
H_{\lambda_{W}\lambda_{D^{*}_{s}}\lambda_{M}}H_{\lambda'_{W}\lambda_{D^{*}_{s}}\lambda_{M}},
\end{align}
where $J$ and $J'$ run over $1$ and $0$, $\lambda^{(')}_{W}$ and $\lambda_{\ell}$ run over their components, $h_{\lambda_{\ell},\lambda_{\bar{\nu}_{\ell}}}=\bar{u}_{\ell}(\lambda_{\ell})\gamma^{\mu}(1-\gamma_{5})\nu_{\bar{\nu}}(\frac{1}{2})\bar{\epsilon}_{\mu}(\lambda_{W})$
is the leptonic helicity amplitude in the $\ell-\bar{\nu}_{\ell}$ center-of-mass frame and can be written as $|h_{-\frac{1}{2},\frac{1}{2}}|^{2}=8(q^{2}-m^{2}_{\ell})$ and
$|h_{\frac{1}{2},\frac{1}{2}}|^{2}=8\frac{m^{2}_{\ell}}{2q^{2}}(q^{2}-m^{2}_{\ell})$. For the standard expression of $d^{J}$ function, we take their value given by PDG~\cite{PDG:2022}.

Finally, we can further evaluate the observables of $D^{*}_{s}\to M\ell\bar{\nu}_{\ell}$ decays. The double differential decay rate is written as
\begin{align}
\frac{\d^{2}\Gamma}{\d q^{2}\d\cos\theta}=\frac{G^{2}_{F}|V_{cs}|^{2}}{(2\pi)^{3}}
\frac{|\vec{p}_{M}|}{8m^{2}_{D^{*}_{s}}}\frac{1}{3}(1-\frac{m^{2}_{\ell}}{q^{2}})L_{\mu\nu}H^{\mu\nu}\,.\label{eq:DG}
\end{align}
Using the amplitudes obtained above, the double differential decay rates of $D^{*}_{s}\to \phi\ell\bar{\nu}_{\ell}$ with a given helicity state of lepton $(\lambda_{\ell}=\pm\frac{1}{2})$ are written as
\begin{align}
\frac{\d^{2}\Gamma[\lambda_{\ell}=1/2]}{\d q^{2}\d\cos{\theta}}
=&\frac{G^{2}_{F}|V_{cs}|^2|\vec{p}_{\phi}|}{256\pi^{3}m^{2}_{D^{*}_{s}}}\frac{1}{3}q^{2}(1-\frac{m^{2}_{\ell}}{q^{2}})^{2}\frac{m^{2}_{\ell}}{q^{2}}
[2(H_{t++}-\cos{\theta}H_{0++})^{2}+2(H_{t--}-\cos{\theta}H_{0--})^{2}\nonumber\\
&+2(H_{t00}-\cos{\theta}H_{000})^{2}+\sin^{2}{\theta}(H^{2}_{+0+}+H^{2}_{+-0}+H^{2}_{-0-}+H^{2}_{-+0})]\,,
\label{eq:dGp1}\\
\frac{\d^{2}\Gamma[\lambda_{\ell}=-1/2]}{\d q^{2}\d\cos{\theta}}
=&\frac{G^{2}_{F}|V_{cs}|^2|\vec{p}_{\phi}|}{256\pi^{3}m^{2}_{D^{*}_{s}}}\frac{1}{3}q^{2}(1-\frac{m^{2}_{\ell}}{q^{2}})^{2}
[2\sin^{2}{\theta}(H_{0++}^{2}+H_{0--}^{2}+H_{000}^{2})\nonumber\\
&+(1-\cos{\theta})^{2}(H_{+-0}^{2}+H_{+0+}^{2})+(1+\cos{\theta})^{2}(H_{-+0}^{2}+H_{-0-}^{2})]\,.
\label{eq:dqmm1}
\end{align}
Integrating over $\cos\theta$ and summing over the lepton helicity, one can obtain the differential decay rate written as
\begin{align}
\frac{\d\Gamma}{\d q^{2}}=&\frac{G^{2}_{F}|V_{cs}|^{2}|\vec{p}_{\phi}|}{96\pi^{3}m^{2}_{D^{*}_{s}}}
\frac{1}{3}q^{2}(1-\frac{m^{2}_{\ell}}{q^{2}})^{2}
\Big[\frac{3m^{2}_{\ell}}{2q^{2}}(H^{2}_{t++}+H^{2}_{t--}+H^{2}_{t00})
+(H^{2}_{+0+}+H^{2}_{+-0}\nonumber\\
&+H^{2}_{-0-}+H^{2}_{-+0}+H^{2}_{000}+H^{2}_{0--}
+H^{2}_{0++})(1+\frac{m^{2}_{\ell}}{2q^{2}})\Big]\,.
\label{eq:DG1}
\end{align}
For the $D^{*}_{s}\to \eta^{(\prime)}\ell\bar{\nu}_{\ell}$ decay, the final state $\eta^{(\prime)}$ is a pseudoscalar meson with spin-zero, and thus carries no hadronic helicity. Consequently, the expressions of double differential decay rates for $\lambda_{\ell}=\pm 1/2$ and  the differential decay rate of $D^{*}_{s}\to \eta^{(\prime)} \ell\bar{\nu}_{\ell}$ decay can be easily obtained from Eqs.~(\ref{eq:dGp1}-\ref{eq:dqmm1}) and Eq.~\eqref{eq:DG1}, respectively,  by neglecting the hadronic helicity amplitudes for $\lambda_\phi=\pm$ and replacing $\phi$ with $\eta^{(\prime)}$. The hadronic helicity amplitudes for $D^{*}_{s}\to \eta^{(\prime)} \ell\bar{\nu}_{\ell}$ decay have been given by Eqs.~(\ref{eq:HAetas1}-\ref{eq:HAetas3}).

Besides the decay rate, there are also two important observables, the lepton spin asymmetry and the forward-backward asymmetry, which are defined as
\begin{align}
&A^{M}_{\lambda}(q^2) \equiv \frac{\d\Gamma[\lambda_{\ell}=-1/2]/\d q^{2}-\d\Gamma[\lambda_{\ell}=1/2]/\d q^{2}}
{\d\Gamma[\lambda_{\ell}=-1/2]/ \d q^{2}+\d\Gamma[\lambda_{\ell}=1/2]/\d q^{2}}\label{eq:Alambda},\\
\intertext{and}
&A^{M}_{\theta}(q^2)\equiv\frac{\int^{0}_{-1}\d\cos\theta(\d^{2}\Gamma/\d q^{2} \d\cos\theta)-\int^{1}_{0}\d\cos\theta(\d^{2}\Gamma/\d q^{2}\d\cos\theta)}
{\d^{2}\Gamma/\d q^{2}}\label{eq:Atheta},
\end{align}
respectively. These observables are independent of the CKM matrix elements, and the hadronic uncertainties canceled to a large extent, therefore, they can be predicted with a rather high accuracy.
\section{Numerical Results and Discussions}
Before presenting our predictions for semileptonic $D^{*}_{s}\to M\ell\bar{\nu}_{\ell}$ and leptonic $D^{*}_{s}\to \ell \bar{\nu}_{\ell}$ weak decays, we would like to clarify the input parameters used in our numerical evaluations.  For the CKM matrix elements, we take $|V_{cs}|=0.987\pm0.011$, which is obtained by averaging the determinations from leptonic $D_s$ and semileptonic $D$ decays~\cite{PDG:2022}.  For the well-known Fermi coupling constant $G_{F}$ and the masses of mesons and lepton, we take the default values given by PDG~\cite{PDG:2022}.
\begin{table}[ht]
\begin{center}
\caption{\label{tab:1}\normalsize  The values of inputs~(in units of GeV) used in the evaluations of decay constants and form factors.}
\vskip 0.25cm
\let\oldarraystretch=\arraystretch
\renewcommand*{\arraystretch}{1.3}
\setlength{\tabcolsep}{13.0pt}
 \begin{tabular}{lc}
        \hline
        \hline
        \text{Parameter} & \text{Value} \\
        \hline
        $m_q$ & $0.230$ \\
        $m_s$ & $0.430$ \\
        $m_c$ & $1.600$ \\
        \hline
        $\beta^{P}_{c\bar q}$       & $0.473 \pm 0.012$ \\
        $\beta^{P}_{c\bar s}$       & $0.531 \pm 0.008$ \\
        $\beta^{V}_{s\bar s}$    & $0.348 \pm 0.006$ \\
        $\beta^{V}_{c\bar q}$    & $0.429 \pm 0.013$ \\
        $\beta^{V}_{c\bar s}$    & $0.496 \pm 0.007$ \\
        \hline
        \hline
    \end{tabular}
\end{center}
\end{table}

\begin{figure}[htbp]
  \centering
  \subfigure[]{\includegraphics[width=0.35\textwidth]{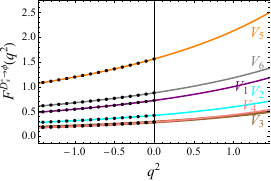}\qquad}
  \subfigure[]{\includegraphics[width=0.35\textwidth]{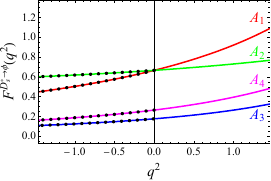}\qquad}
  \subfigure[]{\includegraphics[width=0.35\textwidth]{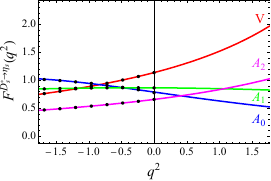}}
\caption{The $q^{2}$-dependences of form factors for $D^{*}_{s}\to \phi$ and $D^{*}_{s}\to \eta_{s}$ transitions.}\label{fig1}
\end{figure}

Besides, the decay constants and form factors are essential nonperturbative  inputs in the calculation.  In this work, we adopt a self-consistent covariant light approach~\cite{V2Vlnu:93,V2Vlnu:94,V2Vlnu:95,Choi:2013mda,Chang:2018zjq}  to evaluate the values of these quantities. The theoretical formulas for the decay constant of vector meson and the form factors of $V'\to V''$ and $V\to P$ transitions have been derived in our previous works~\cite{V'2V'':WangLiTing,P2V:LiXiaoNan} and also are given in the Appendix A. In the evaluation, we shall use the Gaussian-type wavefunction,
\begin{eqnarray}
\psi(x,k_{\bot})=4\frac{\pi^{3/4}}{\beta^{3/2}}
\sqrt{\frac{\partial k_{z}}{\partial x}}{\rm exp}\left[-\frac{k^{2}_{z}+k^{2}_{\bot}}{2\beta^{2}}\right]\,,
\end{eqnarray}
where $\beta$ is the shape parameter, $k_{z}$ is the relative momentum in $z$-direction and has the form $k_{z}=(x-1/2)M_{0}+(m^{2}_{2}-m^{2}_{1})/2M_{0}$.
It is obvious that the quark masses and shape parameters are essential inputs in the numerical evaluation. Their values are collected in Table~\ref{tab:1}.
In practice, the calculation is performed in the Drell-Yan-West frame, $q^{+}=0$, which implies that the form factors are known for space-like momentum transfer, $q^{2}=-q^{2}_{\bot}$, and the ones in the physical time-like region require a $q^{2}$ extrapolation. Following the strategy employed in Refs.~\cite{V2Vlnu:82,V2Vlnu:93,V2Vlnu:94,V2Vlnu:95}, one can parameterize the form factors as functions of $q^{2}$ using modified dipole model in the space-like region and then extend them to the whole physical region,  $0\leq{q^{2}}\leq{(m_{D^{*}_{s}}-m_{M})^{2}}$. The form factors in the modified dipole model have the form
\begin{eqnarray}
F(q^{2})=\frac{F(0)}{1-a(q^{2}/m^{2}_{\rm pole})+b(q^{4}/m^{4}_{\rm pole})}.
\end{eqnarray}
Using the values of input parameters given in Table~\ref{tab:1}, we then present our theoretical predictions for the form factors of $D^{*}_{s}\to \phi$ and $D^{*}_{s}\to \eta_{s}$ transitions in Tables \ref{tab:2} and  \ref{tab:3}, respectively. In addition, the dependences of form factors on  $q^{2}$ are shown in Fig.~\ref{fig1}.
\begin{table}[ht]
\caption{\label{tab:2}Form factors of $D^*_s\to \phi$  in the self-consistent covariant light-front approach.}
\vspace{-0.1cm}\footnotesize
\begin{center}\setlength{\tabcolsep}{8pt}
\begin{tabular}{cccccccccccccccccc}
\hline\hline
                         &$F(0)$                &a   &b
&                          &$F(0)$                &a   &b     \\  \hline
  $V_1^{D^*_s\to \phi}$     &$0.72^{+0.01}_{-0.01}$&$1.35^{+0.00}_{-0.00}$ &$0.47^{+0.00}_{-0.00}$
& $V_2^{D^*_s\to \phi}$     &$0.42^{+0.02}_{-0.02}$&$1.41^{+0.01}_{-0.02}$ &$0.47^{+0.01}_{-0.01}$ \\
  $V_3^{D^*_s\to \phi}$     &$0.28^{+0.00}_{-0.00}$&$1.52^{+0.02}_{-0.02}$ &$0.66^{+0.03}_{-0.03}$
& $V_4^{D^*_s\to \phi}$     &$0.28^{+0.00}_{-0.00}$&$1.70^{+0.02}_{-0.02}$&$0.81^{+0.03}_{-0.03}$ \\
  $V_5^{D^*_s\to \phi}$     &$1.56^{+0.02}_{-0.02}$&$1.28^{+0.00}_{-0.00}$&$0.39^{+0.00}_{-0.00}$
& $V_6^{D^*_s\to \phi}$     &$0.87^{+0.01}_{-0.01}$&$1.24^{+0.01}_{-0.01}$&$0.36^{+0.00}_{-0.00}$ \\
  $A_1^{D^*_s\to \phi}$     &$0.66^{+0.01}_{-0.01}$&$1.34^{+0.00}_{-0.00}$&$0.46^{+0.00}_{-0.00}$
& $A_2^{D^*_s\to \phi}$     &$0.66^{+0.01}_{-0.01}$&$0.37^{+0.03}_{-0.03}$&$-0.12^{+0.01}_{-0.01}$ \\
  $A_3^{D^*_s\to \phi}$     &$0.17^{+0.01}_{-0.01}$&$0.81^{+0.03}_{-0.03}$&$1.67^{+0.03}_{-0.03}$
& $A_4^{D^*_s\to \phi}$     &$0.26^{+0.00}_{-0.00}$&$1.66^{+0.02}_{-0.02}$&$0.84^{+0.03}_{-0.03}$ \\
  \hline\hline
\end{tabular}
\end{center}
\end{table}

\begin{table}[ht]
\caption{\label{tab:3}Form factors of $D^*_s\to \eta_{s}$ in the self-consistent covariant light-front approach.}
\vspace{-0.1cm}\footnotesize
\begin{center}\setlength{\tabcolsep}{8pt}
\begin{tabular}{cccccccccccccccccc}
\hline\hline
 F                         &$F(0)$                &a   &b \\  \hline
$V^{D^*_s\to \eta_{s}}$      &$1.13^{+0.01}_{-0.00}$&$1.19^{+0.00}_{-0.00}$ &$0.32^{+0.00}_{-0.01}$ \\
$A_0^{D^*_s\to \eta_{s}}$    &$0.78^{+0.01}_{-0.00}$&$-0.81^{+0.00}_{-0.00}$&$0.67^{+0.00}_{-0.00}$ \\
$A_1^{D^*_s\to \eta_{s}}$    &$0.86^{+0.00}_{-0.00}$&$-0.04^{+0.00}_{-0.00}$&$0.43^{+0.00}_{-0.00}$ \\
$A_2^{D^*_s\to \eta_{s}}$    &$0.66^{+0.01}_{-0.02}$&$1.34^{+0.01}_{-0.01}$&$0.37^{+0.01}_{-0.01}$ \\
  \hline\hline
\end{tabular}
\end{center}
\end{table}

\begin{figure}[htbp]
  \centering
  \subfigure[]{\includegraphics[width=0.38\textwidth]{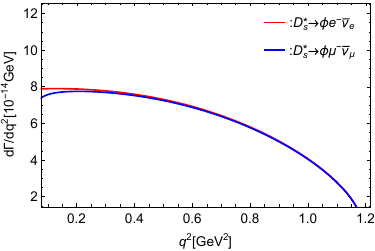}\quad\qquad}
  \subfigure[]{\includegraphics[width=0.38\textwidth]{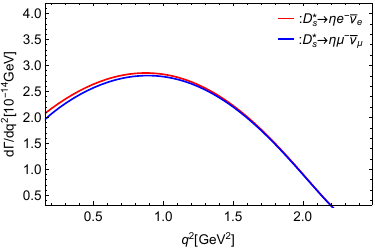}\quad\qquad}
  \subfigure[]{\includegraphics[width=0.38\textwidth]{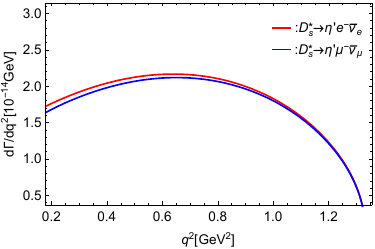}\quad\qquad}
\caption{The $q^{2}$-dependences of differential decay rates $\d \Gamma/\d q^{2}$.}\label{fig2}
\end{figure}
\par
Based on the theoretical formulas in the previous section and the form factors obtained in the self-consistent covariant light-front quark model, the decay widths of the leptonic $D^{*}_{s}\to\ell\bar{\nu}_{\ell}$ and semileptonic $D^{*}_{s}\to(\phi,\eta^{(\prime)})\ell\bar{\nu}_{\ell}$ weak decays can be obtained and are listed in Table \ref{tab:4}. Furthermore, the $q^{2}$-dependence of the differential decay rate ${\d}\Gamma/{\d}q^{2}$ is shown in Fig.~\ref{fig2}. For the decay widths of $D^{*}_{s}\to (\phi, \eta^{(\prime)})\ell\bar{\nu}_{\ell}$, the two errors arise from the uncertainties of the form factors and the CKM matrix elements $V_{cs}$, respectively. Similarly, the two errors in the decay width of leptonic $D^{*}_{s}\to\ell\bar{\nu}_{\ell}$ weak decay are caused by the uncertainties of the decay constant $f_{D^{*}_{s}}$ and the CKM matrix elements $V_{cs}$, respectively.
\begin{table}[ht]
\begin{center}
\caption{\label{tab:4} \small The theoretical predictions for the decay width of $ D^{*}_{s} \to (\phi,\eta^{(\prime)})\ell\bar{\nu}_{\ell}$ and $D^{*}_{s}\to\ell\bar{\nu}_{\ell}$ decays.}
\vspace{0.2cm}
\let\oldarraystretch=\arraystretch
\renewcommand*{\arraystretch}{1.3}
\setlength{\tabcolsep}{13.0pt}
\begin{tabular}{lcccc}
\hline\hline
  Decay mode
  &decay width    \\
  \hline
  $D^{*-}_{s} \to \phi e^{-}\bar{\nu}_{e}$
  &$7.38^{+0.25+0.16}_{-0.25-0.16}\times10^{-14}$
  \\
  $D^{*-}_{s} \to \phi\mu^{-}\bar{\nu}_{\mu}$
  &$7.08^{+0.22+0.16}_{-0.22-0.16}\times10^{-14}$
  \\
  $D^{*-}_{s} \to \eta e^{-}\bar{\nu}_{e}$
  &$5.09^{+0.02+0.11}_{-0.02-0.11}\times10^{-14}$
  \\
  $D^{*-}_{s} \to \eta\mu^{-}\bar{\nu}_{\mu}$
  &$4.96^{+0.02+0.11}_{-0.02-0.11}\times10^{-14}$
  \\
  $D^{*-}_{s} \to \eta^{\prime} e^{-}\bar{\nu}_{e}$
  &$2.37^{+0.01+0.05}_{-0.01-0.05}\times10^{-14}$
  \\
  $D^{*-}_{s} \to \eta^{\prime}\mu^{-}\bar{\nu}_{\mu}$
  &$2.26^{+0.01+0.05}_{-0.01-0.05}\times10^{-14}$
  \\\hline
  $D^{*}_{s} \to \ell\bar{\nu}_{\ell}$
  &$2.56^{+0.09+0.06}_{-0.09-0.06}\times10^{-12}$\\
\hline\hline
\end{tabular}
\end{center}
\end{table}

In order to evaluate the branching fractions of semileptonic $D^{*}_{s}\to M\ell\bar{\nu}_{\ell}$ and leptonic $D^{*}_{s}\to \ell \bar{\nu}_{\ell}$ decays, the total decay width $\Gamma_{\rm tot}^{D^{*}_{s}}$ is the essential input. However, there is no available experimental information for  $\Gamma_{\rm tot}^{D^{*}_{s}}$ until now.
Fortunately,  the branching fraction of $D^{*}_{s} \to D_{s}\gamma$ radiation decay has been well measured~\cite{PDG:2022},
\begin{eqnarray}
{\cal B}(D^{*+}_{s}\to D^{+}_{s}\gamma)_{\rm{exp.}}=(93.6\pm 0.4)\%\,,
\end{eqnarray}
and the full width for $D^{*}_s$ meson can be extracted through the relation
\begin{eqnarray}
\Gamma^{D^*_{s}}_{\rm tot}=\frac{\Gamma(D^*_{s}\to D_{s}\gamma)}{\mathcal{B}(D^*_{s}\to D_{s}\gamma)}\label{LFFG}\,.
\end{eqnarray}
The decay width $\Gamma(D^*_{s}\to D_{s}\gamma)$ can be calculated by using the self-consistent covariant light front approach, and is written as~\cite{V2Vlnu:86},
\begin{eqnarray}
\Gamma(D^{*}_{s}\to D_{s}\gamma)_{\rm LF}&=&\frac{\alpha}{3}[e_{1}I(m_{1},m_{2},0)
+e_{2}I(m_{2},m_{1},0)]^{2}\kappa^{3}_{\gamma}\,,\\
I(m_{1},m_{2},q^{2})&=&\int^{1}_{0}\frac{dx}{8\pi^{3}}
\int d^{2}k_{\bot}\frac{\psi(x,k'_{\bot})\psi(x,k_{\bot})}
{x\tilde{M}_{0}\tilde{M'}_{0}}\nonumber\\
&&\times
\left\{[(1-x)m_{1}+xm_{2}]+\frac{2}{M_{0}+m_{1}+m_{2}}\left[
k^{2}_{\bot}-\frac{(k_{\bot}\cdot q_{\bot})^{2}}{q^{2}_{\bot}}\right]\right\}\,,
\end{eqnarray}
where $M_{0}$ is the invariant mass and $\kappa_{\gamma}=(m_{D^{*}_{s}}^{2}-m^{2}_{D_{s}})^{2}/2m_{D^{*}_{s}}$ is the kinematically allowed energy of the outgoing photon.
Using the values of input parameters given in Table~\ref{tab:1}, we obtain
\begin{eqnarray}
\Gamma(D^{*}_{s}\to D_{s}\gamma)_{\rm LF}=0.123^{+0.003}_{-0.003}\, \rm{KeV}\,,\label{FG}
\end{eqnarray}
which is generally in agreement with the ones obtained in the previous work~\cite{V2Vlnu:86,V2Vlnu:87,V2Vlnu:88,V2Vlnu:89,V2Vlnu:90,V2Vlnu:91,V2Vlnu:92}. Employing Eqs.(\ref{LFFG})  and (\ref{FG}), one can finally obtain the full width of $D^{*}_{s}$ meson,
\begin{eqnarray}\label{G1}
\Gamma^{D^{*}_{s}}_{\rm tot}=0.131^{+0.003}_{-0.003}\,\rm{KeV}\,.
\end{eqnarray}
\par
\begin{table}[ht]
\scriptsize
\begin{center}
\caption{\label{tab:5} \small The theoretical predictions for the branching fractions of $D^{*}_{s} \to (\phi,\eta^{(\prime)})\ell\bar{\nu}_{\ell}$ and $D^{*}_{s}\to\ell\bar{\nu}_{\ell}$ decays.}
\vspace{0.2cm}
\let\oldarraystretch=\arraystretch
\renewcommand*{\arraystretch}{1.3}
\setlength{\tabcolsep}{3.8pt}
\begin{tabular}{lcccccccccc}
\hline\hline
  Decay mode
  &this work & \cite{Dstar2phi:ChengShan}   &~\cite{Yang:2021crs,Yang:2025hzk}
  &BESIII~\cite{BESIII:2023zjq} \\
  \hline
  $ D^{*-}_{s} \to \phi e^{-}\bar{\nu}_{e}$
  &$(5.63\pm{0.19\pm0.13\pm0.13})\times10^{-7}$ &$(0.47^{+0.12}_{-0.10}\pm0.19)\times10^{-6}$
  &
  &-
  \\
  $D^{*-}_{s} \to \phi\mu^{-}\bar{\nu}_{\mu}$
  &$(5.40\pm{0.17\pm0.12\pm0.12})\times10^{-7}$ &$(0.47^{+0.12}_{-0.10}\pm0.19)\times10^{-6}$
  &
  &-
  \\
  $D^{*-}_{s} \to \eta e^{-}\bar{\nu}_{e}$
  &$(3.89\pm{0.02\pm0.09\pm0.09})\times10^{-7}$
  &
  &$(1.46^{+0.16+0.03}_{-0.12-0.09})\times10^{-6}$~\cite{Yang:2025hzk}
  &-
  \\
  $D^{*-}_{s} \to \eta \mu^{-}\bar{\nu}_{\mu}$
  &$(3.79\pm{0.02\pm0.09\pm0.09})\times10^{-7}$
  &
  &$(1.41^{+0.15+0.02}_{-0.12-0.09})\times10^{-6}$~\cite{Yang:2025hzk}
  &-
  \\
  $D^{*-}_{s} \to \eta'e^{-}\bar{\nu}_{e}$
  &$(1.81\pm{0.01\pm0.04\pm0.04})\times10^{-7}$
  &
  &$(5.08^{+0.55+0.10}_{-0.42-0.37})\times10^{-7}$~\cite{Yang:2025hzk}
  &-
  \\
  $D^{*-}_{s} \to \eta'\mu^{-}\bar{\nu}_{\mu}$
  &$(1.73\pm{0.01\pm0.04\pm0.04})\times10^{-7}$
  &
  &$(4.80^{+0.52+0.10}_{-0.40-0.36})\times10^{-7}$~\cite{Yang:2025hzk}
  &-
  \\\hline
  $D^{*}_{s} \to \ell\bar{\nu}_{\ell}$
  &$(1.95\pm{0.07\pm0.04\pm0.04})\times10^{-5}$
  &$(3.49\pm0.14)\times10^{-5}$
  &$(6.7\pm0.4)\times 10^{-5}$~\cite{Yang:2021crs}
  &$(2.1^{+1.2}_{-0.9}\pm0.2)\times 10^{-5}$\\
\hline\hline
\end{tabular}
\end{center}
\end{table}

Based on the theoretical formulas provided above, we present our numerical results for the branching ratios of the leptonic $D^{*}_{s}\to \ell\bar{\nu}_{\ell}$ and semileptonic $D^{*}_{s}\to (\phi,\eta^{(\prime)})\ell\bar{\nu}_{\ell}$ weak decays  in Table \ref{tab:5}. For the convenience of comparison, we have also summarized the relevant results obtained from other works in Table \ref{tab:5}\cite{BESIII:2023zjq,Dstar2phi:ChengShan,Yang:2021crs,Yang:2025hzk}.
There are three theoretical errors in our numerical results, which originate from the uncertainty of form factors, the CKM matrix elements $V_{cs}$, and the total width of $D^{*}_{s}$ meson, respectively. Numerically, the semileptonic $D^{*}_{s}\to (\phi,\eta^{(\prime)})\ell\bar{\nu}_{\ell}$ weak decays have large branching ratios, which can reach up to the order of $10^{-7}$. Hence, they are most likely to be detected by future
high-luminosity experiments, such as the SuperKEKB, STCF, and LHCb.
\par
As shown in Table \ref{tab:5}, for the leptonic $D^{*}_{s}\to \ell\bar{\nu}_{\ell}$ weak decay, our result is in good agreement with the experimental data of BESIII collaboration within the uncertainty. For the decay mode of $D^{*}_{s}\to \phi\ell\bar{\nu}_{\ell}$, our results are consistent with other theoretical results~\cite{Dstar2phi:ChengShan} when the uncertainties of all calculations are considered. However, the branching ratios of semileptonic $D^{*}_{s}\to \eta \ell\bar{\nu}_{\ell}$ and $D^{*}_{s}\to \eta^{\prime} \ell\bar{\nu}_{\ell}$ decays are significantly lower than the results presented in Ref.\cite{Yang:2025hzk}. This observed discrepancy originates from different treatments of the $\eta$-$\eta'$ mixing mechanism and the relevant form factors. Future experimental measurements will be crucial to resolving these theoretical discrepancies.
\begin{figure}[htbp]
  \centering
  \subfigure[]{\includegraphics[width=0.35\textwidth]{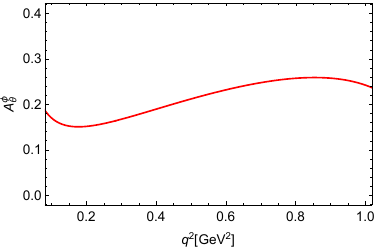}\quad\qquad}
  \subfigure[]{\includegraphics[width=0.35\textwidth]{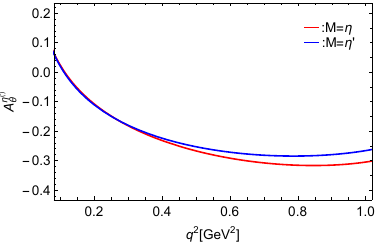}\quad\qquad}
  \subfigure[]{\includegraphics[width=0.35\textwidth]{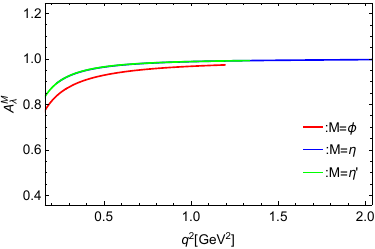}\quad\qquad}
\caption{The $q^{2}$-dependences of $A^{\phi,\eta^{(\prime)}}_{\theta}$ and $A^{\phi,\eta^{(\prime)}}_{\lambda}$.}\label{fig3}
\end{figure}

\begin{table}[ht]
\begin{center}
\caption{\label{tab:6} \small Predictions for $q^{2}$-integrated observables $A^{\phi,\eta^{(\prime)}}_{\theta,\lambda}$.}
\vspace{0.2cm}
\let\oldarraystretch=\arraystretch
\renewcommand*{\arraystretch}{1.1}
\setlength{\tabcolsep}{8.8pt}
\begin{tabular}{lcccccccccc}
\hline\hline
  Observable  &&Prediction\\
  \hline
  \qquad$A^{\phi}_{\lambda}$   &&$0.868^{+0.008}_{-0.008}$\\
  \qquad$A^{\eta}_{\lambda}$   &&$0.961^{+0.003}_{-0.002}$\\
  \qquad$A^{\eta^{\prime}}_{\lambda}$   &&$0.937^{+0.002}_{-0.001}$\\\hline
  \qquad$A^{\phi}_{\theta}$   &&$0.172^{+0.004}_{-0.004}$\\
  \qquad$A^{\eta}_{\theta}$  &&$-0.361^{+0.005}_{-0.003}$\\
  \qquad$A^{\eta^{\prime}}_{\theta}$  &&$-0.213^{+0.002}_{-0.001}$\\\hline
  \hline
\end{tabular}
\end{center}
\end{table}
\par
We present the numerical results for the lepton spin asymmetry $A^{\phi,\eta^{(\prime)}}_{\theta}$ and the forward-backward asymmetry $A^{\phi,\eta^{(\prime)}}_{\lambda}$ defined by Eqs.(\ref{eq:Alambda}) and (\ref{eq:Atheta}) in Table \ref{tab:6}. The theoretical uncertainties are caused only by the form factors. It can be seen from Eq.~(\ref{eq:Alambda}) and (\ref{eq:Atheta}), the hadronic uncertainties are significantly reduced due to cancellations between the numerators and denominators in the definitions of these asymmetries. Consequently, their relative uncertainties are much smaller than those of the branching fractions. Futhermore, we display the $q^2$-dependences of the lepton spin asymmetry and the forwardbackward asymmetry in Fig.~\ref{fig3}.
As shown in Table \ref{tab:6} and Fig.\ref{fig3}(c), we observe that the values of the forward-backward asymmetries $A^{\phi}_{\lambda}$ and $A^{\eta^{(\prime)}}_{\lambda}$ are almost equal to each other. For the differential distributions of the lepton spin asymmetries $A^{\eta}_{\theta}$ and $A^{\eta^{\prime}}_{\theta}$, a characteristic feature is the zero-crossing point, which is usually used to distinguish the new physics (NP) effects from the SM, or different NP scenarios. Additionally, as shown in Fig.\ref{fig3}(b), one can easily find that $A^{\eta^{(\prime)}}_{\theta}$ cross the zero point at $q^{2}\approx 0.07\rm{GeV}^{2}$. These physical observables play an important role in testing the SM of particle physics.
\section{Summary}
\par
Inspired by the recent advancements and future prospects for the study of $D^{*}_{s}$ mesons in collider experiments, we systematically investigate the leptonic $D^{*}_{s}\to\ell\bar{\nu}_{\ell}$ and  semileptonic $D^{*}_{s}\to(\phi,\eta^{(\prime)})\ell\bar{\nu}_{\ell}$ weak decays. We obtained the decay constant and form factors of $D^{*}_{s}$ meson, and the decay width $\Gamma(D^{*}_{s}\to D_{s}\gamma)$ by using the self-consistent covariant light front approach. The decay widths, branching ratios, the lepton spin asymmetry and the forward-backward asymmetry of the corresponding $D^{*}_{s}$ weak decays are calculated. Our numerical results show that the CKM-favored $D^{*}_{s}\to (\phi, \eta^{(\prime)})\ell\bar{\nu}_{\ell}$ decays have relatively large branching fractions of $\mathcal{O}(10^{-7})$, and hence are expected to be measured by the heavy-flavor experiments at the SuperKEKB/Belle-II, STCF, and LHCb.
\section*{Acknowledgments}
This work is supported by the National Natural Science Foundation of China (Grants Nos.12275067 and 12305101);
Science and Technology R$\&$D Program Joint Fund Project of Henan Province  (Grant No.225200810030);
Science and Technology Innovation Leading Talent Support Program of Henan Province  (Grant No. 254200510039);
Key Research Project Plan for Higher Education Institutions of Henan Province (Grant No.23A140012).
\section*{Appendix A}
The $D^{*}_{s}\to M$ transition form factors, defined in Eqs.(\ref{eq:FF1}-\ref{eq:FF4}), are also crucial inputs for evaluating physical observables, especially for the branching fraction. In this work, we adopt self-consistent covariant light front approach to evaluate their values\cite{V2Vlnu:93,V2Vlnu:94,V2Vlnu:95}. The theoretical formulas for the form factors of $V'\to V''$ have been given in our previous work (see Eqs.(3.23-3.33) in Ref.\cite{V'2V'':WangLiTing}). The form factors of $V\to P$ transition are given by\cite{P2V:LiXiaoNan}
\begin{eqnarray}\label{eq:FCLF}
[\mathcal F(q^2)]_{\rm CLF}=N_c\int\frac{dxd^2{\bf k'_\bot}}{2(2\pi)^3}\frac{\chi_M'\chi_M''}{\bar x}{\cal \widetilde{\cal F}}^{\rm CLF}(x,{\bf k'_\bot},q^2)\,,
\end{eqnarray}
where the integrands are
\begin{align}
\label{eq:VCLF}
[f]^{\rm CLF}=&-2\bigg\{m'_1\Big[M''^2-(m''_1-m_2)^2-q^2-\hat N'_1-\hat Z_2\Big]\nonumber\\
&+m''_1(M'^2-m'^2_1-m_2^2-\hat N'_1-\hat Z_2 +4A^{(2)}_1)\nonumber\\
&+2m_2( m'^2_1+\hat N'_1-2A^{(2)}_1)+\frac{2}{D_{V,{\rm con}}}\Big[(2x-1)M'^2+M''^2-2(m'_1-m''_1)(m'_1+m_2)\nonumber\\
&+2xM'^2_0-q^2-\frac{2\kb\cdot\qb}{q^2}(M'^2-M''^2+q^2)\Big]\Big[ \kb^2+ \frac{(\kb\cdot\qb)^2}{q^2}
\Big]
\bigg\}\,,
\end{align}
\begin{align}
[a_+]^{\rm CLF}=&2m'_1(1-A^{(1)}_1-A^{(1)}_2)-\frac{4\kb''\cdot\qb}{\bar x q^2 D_{V,{\rm con}}}\Big[\kb\cdot\kb''+(\bar xm''_1+xm_2)(xm_2-\bar xm'_1)\Big]\nonumber\\
&+2m''_1(A^{(1)}_1-A^{(1)}_2-4A^{(2)}_2+4A^{(2)}_3)+4m_2(A^{(1)}_1+2A^{(2)}_2-2A^{(2)}_3)
\,,
\end{align}

\begin{align}
[a_-]^{\rm CLF}=&2\bigg\{m'_1(A^{(1)}_1+A^{(1)}_2+1)+m''_1(A^{(1)}_2-A^{(1)}_1+4A^{(2)}_3-4A^{(2)}_4)\nonumber\\
&-\frac{1}{D_{V,{\rm con}}}\Big[ -2(M'^2+M''^2+2(m'_1+m_2)(m''_1-m_2)-q^2)(A^{(2)}_3+A^{(2)}_4-A^{(1)}_2)\nonumber\\
&+\Big(2M'^2+(m'_1+m''_1)^2-2(m'_1+m_2)^2-q^2-\hat N'_1+\hat N''_1\Big)(A^{(1)}_1+A^{(1)}_2-1)\nonumber\\
&+2\hat Z_2(2A^{(2)}_4-3A^{(1)}_2+1)
+2\frac{M'^2-M''^2}{q^2}(4A^{(1)}_2A^{(2)}_1-3A^{(2)}_1)
\Big]\nonumber\\
&+2m_2(A^{(1)}_1-2A^{(1)}_2-2A^{(2)}_3
+2A^{(2)}_4)
\bigg\}\,,\\
[g]^{\rm{CLF}}=&-2\bigg\{\bar x m'_1+xm_2+(m'_1-m''_1)\frac{\kb\cdot\qb}{q^2}+\frac{2}{D_{V,{\rm con}}}\left[ \kb'^2+ \frac{(\kb'\cdot\qb)^2}{q^2}
\right]\bigg\}\,,
\end{align}

\begin{align}
&V(q^2)= -(M'+M'')g(q^2)\,,\\
&A_0(q^2)=-\frac{1}{2M'}\left[ -(M'^2-M''^2)a_+(q^2)+f(q^2)+q^2a_-(q^2)\right]\,,\\
&A_1(q^2)=-\frac{f(q^2)}{M'+M''}\,,\\
&A_2(q^2)=(M'+M'')a_+(q^2)\,.
\end{align}

\end{document}